\documentclass[12pt,aps,prd,preprint,tightenlines,superscriptaddress, amsmath,amssymb, nofootinbib]{revtex4}
\pdfoutput=1
\usepackage{amsmath,graphics,epsfig}
\usepackage{color,hyperref}
\usepackage{datetime}
\usepackage{booktabs}
\begin{document}

\title{Light Neutralino Dark Matter in the NMSSM}
\author{Qiurong Mou}
\email{qiurongmou@163.com}
\affiliation{\small{Department of Physics, Chongqing University, Chongqing 401331, P. R. China}}
\author{Hongyan Wu}
\email{hongyanwu@cqu.edu.cn}
\affiliation{\small{Department of Physics, Chongqing University, Chongqing 401331, P. R. China}}
\author{Sibo Zheng}
\email{sibozheng.zju@gmail.com}
\affiliation{\small{Department of Physics, Chongqing University, Chongqing 401331, P. R. China}}
\date{February 21, 2018}

\begin{abstract}
The next-to-minimal supersymmetric standard model can host light neutralino dark matter with mass of order GeV scale.
It is dominated by the singlino component as a result of approximate Peccei-Quinn symmetry.
This paper is devoted to address the question how light such neutralino dark matter can be in the light of the LHC Run 1 data as well as the latest LUX and Xenon1T limits.
In particular, we show the sensitivity of parameter space of dark matter mass
with respect to $Z$ boson and SM Higgs invisible decay.
\end{abstract}

\maketitle
\setcounter{page}{0}
\thispagestyle{empty}

\newpage
\section{Introduction}
It is well known that the Standard Model (SM)-like Higgs discovered at the LHC \cite{HiggsMass1,HiggsMass2} 
imposes severe constraint on stop soft mass parameters.
They are at least several TeVs in order to provide significant radioactive correction to Higgs mass \cite{MSSMHiggs1,MSSMHiggs2,MSSMHiggs3}
in the situation of small mixing or of small tree-level modification.
Similar favor to large soft masses for the others such as gluino mass is also suggested by the LHC Run 1 data.

When the soft mass parameters of supersymmetry are large (in compared with the weak scale), 
it is hardly available to directly detect them at the LHC,
unless some of dimensionless parameters of supersymmetry are extremely large or small.
Such specific choices reopen the potential of leaving striking signatures in collider or astrophysical experimental facilities.

In supersymmetric models favored by simplicity, 
the Minimal Supersymmetric Standard Model (MSSM) and the Next-to-Minimal Supersymmetric Standard Model (NMSSM) (For a review, see, e.g.,  \cite{Review})
are the two simplest ones which retain the unification \cite{1706.01071}  of SM gauge coupling constants.
The later one is an extension of the former by adding a singlet $S$ with superpotential 
\begin{eqnarray}{\label{Superpotential}}
W=\lambda SH_{u}H_{d}+\frac{\kappa}{3}S^{3}.
\end{eqnarray}
The dimensionless parameters in the NMSSM are thus composed of $\lambda$ and $\kappa$ of Eq.(\ref{Superpotential}). 
Following previous line it is prior to tune $\lambda$ or $\kappa$ in the sense that 
it may leave detectable signatures and help us distinguish NMSSM from MSSM simultaneously.

Remarkably, a small value of $\kappa$ yields a singlino-like neutralino dark matter (DM) with mass as light as a few GeVs.
The smallness of $\kappa$ can be simply analyzed as an input parameter as in earlier studies 
\cite{Singlino2,Singlino3, Singlino4,1510.00246, Singlino5,Singlino6, 1311.0678, Singlino7,Singlino8,Singlino9,Singlino10,Singlino11}.
Alternatively,  it is a result of approximate Peccei-Quinn (PQ) symmetry
\footnote{If $\kappa$ term vanishes the Lagrangian would be invariant under the following $U(1)$ symmetry transformation,
\begin{eqnarray}{\label{PQ}}
H_{u}\rightarrow H_{u} \exp({i\phi}), ~~~~H_{d}\rightarrow H_{d} \exp({i\phi}), ~~~~
S\rightarrow S \exp({-2i\phi}). \nonumber
\end{eqnarray}} as we choose here.
Unlike in the large mass region, 
a light DM is rather sensitive to $Z$ boson \cite{PQNMSSM1,PQNMSSM2, PQNMSSM3} 
and Higgs \cite{1112.1014,1208.2555} invisible decay limits.
Consider that the later one will be significantly improved in comparison with earlier studies,
it is meaningful to uncover the sensitivity of parameter space of DM mass to this experimental value. 

The plan of this paper is organized as follows.
In Sec.2 we firstly analyze the mass matrixes and interactions under the approximate PQ symmetry both in the Higgs and neutralino sector.
Then we use the numerical code NMSSMTools 5.0.2 \cite{NMSSMTool} to solve eigenstate masses, 
and micrOMEGAs \cite{1407.6129} to extract parameter space imposed by DM relic abundance 
as well as direct detection limits at LUX \cite{LUX2016, LUXSD} and Xenon1T \cite{Xenon1T, XenonSD}.
Sec.3 is devoted to DM indirect detections at particle colliders.
We show the constraints arising from $Z$ boson \cite{0509008} and latest Higgs \cite{1606.02266} invisible decay at the LHC.
Finally, we conclude in Sec.4.  
The appendix is added to introduce our convention and notations.

\section{Light Singlino Dark Matter in NMSSM}
The neutralino DM is dominated by the singlino component in PQ-symmetric NMSSM,
as inferred from the neutralino mass matrix in Eq.(\ref{NeutralinoMass}).
The DM mass eigenvalue is generally handled by numerical calculation.
In contrast, analytic approximations can be only obtained under the decoupling limit  $M_{1,2}>>\mu$,
which are useful for illustrating the numerical results in the text.
When the wino and bino are decoupled in Eq.(\ref{NeutralinoMass}),
the remaining three mass eigenstates ordered in mass are then decomposed as,
\begin{eqnarray}{\label{Decomposition}}
\tilde{\chi}^{0}_{i}\simeq N_{i3} \tilde{H}_{u}+N_{i4}\tilde{H}_{d}+N_{i5}\tilde{s},
\end{eqnarray}
where $N$ is a unitary matrix to diagonalize the remaining neutralino mass $M_{\chi}$.
The matrix elements in Eq.(\ref{Decomposition}) are approximated by \cite{Approximation},
\begin{eqnarray}{\label{Ns}}
N_{i3}:N_{i4}:N_{i5}\simeq \lambda(\mu\upsilon_{u}-\upsilon_{d} m_{\tilde{\chi}^{0}_{i}}):\lambda(\mu\upsilon_{d}-\upsilon_{u} m_{\tilde{\chi}^{0}_{i}})
:(m^{2}_{\tilde{\chi}^{0}_{i}}-\mu^{2}),
\end{eqnarray}
where $m_{\tilde{\chi}^{0}_{i}}$ refers to the mass of $\tilde{\chi}^{0}_{i}$.

\subsection{Relic Abundance}
The relic abundance of thermal DM is determined by the averaged annihilation cross section $\left<\sigma v\right>$ 
at the freeze-out time, which is mainly given by the annihilation of singlino to $\tau\tau$ or $b\bar{b}$ through the light CP odd scalar $A_{1}$ if kinetically allowed.
As shown in \cite{Singlino9} it is approximated as,
\begin{eqnarray}{\label{Annihilation}}
\left<\sigma v\right>&\simeq& \frac{g^{2}_{2}c_{f}}{8\pi}\frac{m^{2}_{f}}{M^{2}_{W}}\cos^{2}\theta_{A}\tan^{2}\beta\times \mid T_{A_{1}\tilde{\chi}^{0}_{1}\tilde{\chi}^{0}_{1}}\mid^{2}
\times \frac{m^{2}_{\tilde{\chi}^{0}_{1}}\sqrt{1-m^{2}_{f}/m^{2}_{\tilde{\chi}^{0}_{1}}}}{(4m^{2}_{\tilde{\chi}^{0}_{1}}-m^{2}_{A_{1}})^{2}+m^{2}_{A_{1}}\Gamma^{2}_{A_{1}}},
\end{eqnarray}
where $c_{f}=1(3)$ for lepton (quark), $m_{f}$ is the SM fermion mass, 
$T_{A_{1}\tilde{\chi}^{0}_{1}\tilde{\chi}^{0}_{1}}$ denotes the the CP-odd Higgs-neutralino-neutralino coupling,
$m_{A_{1}}$ is the lighter CP-odd scalar mass, 
and $\cos\theta_{A}$ refers to the mixing angle between the CP-odd scalar $A$ of MSSM and $S_{I}$ of the singlet.
The magnitude of $T_{A_{1}\tilde{\chi}^{0}_{1}\tilde{\chi}^{0}_{1}}$ reads as 
\begin{eqnarray}{\label{Tacc}}
 T_{A_{1}\tilde{\chi}^{0}_{1}\tilde{\chi}^{0}_{1}}\simeq \sqrt{2}\lambda\cos\theta_{A}N_{15}(N_{13}\sin\beta+N_{14}\cos\beta)+\sqrt{2}\sin\theta_{A}(\lambda N_{13}N_{14}-\kappa N^{2}_{15}).
\end{eqnarray}
We refer the reader to the appendix for the other parameters in Eq.(\ref{Annihilation}).
One obtains the estimate of relic abundance by substituting Eq.(\ref{Annihilation}) into the standard formula,
\begin{eqnarray}{\label{RB}}
\Omega_{\tilde{\chi}^{0}_{1}}h^{2}\simeq \frac{10^{9}\text{GeV}^{-1}}{M_{P}}\frac{x_{F}}{\sqrt{g_{*}}}\frac{1}{\left<\sigma v\right>},
\end{eqnarray}
where $M_{P}$ is the Planck mass, $x_{F}\sim 20$,
and $g_{*}$ is effective number of freedoms at the freeze-out temperature.
The experimental value of DM relic abundance reported by Planck and WMAP \cite{Planck} is given as
\begin{eqnarray}{\label{Planck}}
\Omega_{\tilde{\chi}^{0}_{1}}h^{2}=0.1197\pm 0.0022.
\end{eqnarray}

\begin{table}[htb]
\begin{center}
\begin{tabular}{c} 
$\text{Parameter~range}$ 
\\
\hline\hline
 $ 0.1\leq \lambda \leq 0.7$   \\
 $5< \tan\beta \leq 30 $    \\
  $0.005\leq \kappa \leq 0.07$    \\
  $103 \leq \mid \mu \mid \leq 500$    \\
  $-2000 \leq A_{\lambda}\leq 2000$  \\
  $-500 \leq A_{\kappa}\leq 500$  \\
  \hline\hline
  $100\leq M_{1,2} \leq 1000$    \\
  $1000 \leq M_{3} \leq 2500$    \\
  $-3000 \leq A_{t,b}\leq 3000$  \\
  $800 \leq m_{\tilde{t},\tilde{b}}\leq 2000$  \\
\hline\hline
\end{tabular}
\caption{Scanned parameter ranges, where soft mass parameters are in unit of GeV.
Large gluino and stop masses are chosen by referring their lower bounds at the LHC
($\sim 1.5$ TeV and $\sim 740$ GeV for gluino and stop, respectively \cite{1701.01954}) 
whereas  small $\kappa$ is adopted in the spirit of spontaneously broken PQ symmetry.}
\label{data}
\end{center}
\end{table}

Obviously, the parameter space induced by DM relic density is sensitive to the two mixing angles and masses of DM and light CP-odd scalar in Eq.(\ref{Annihilation}),
which are directly related to parameters $\lambda$,  
$\tan\beta$, $\kappa$, $\mu$, $A_{\lambda}$ and $A_{\kappa}$ as shown in the appendix,
and indirectly to parameters such as gaugino masses $M_{i}$ as well as the soft masses involving the third generation.
The parameter ranges for them to be scanned in terms of the numerical codes NMSSMTools 5.0.2 \cite{NMSSMTool} and micrOMEGAs \cite{1407.6129} are explicitly shown in Table.\ref{data}.
What is the most different in our scans of parameter ranges from the other studies in the literature is that
large gluino and stop masses are chosen whereas small $\kappa$ is adopted in the spirit of spontaneously broken PQ symmetry.

\begin{figure}
\centering
\includegraphics[width=7cm,height=8cm]{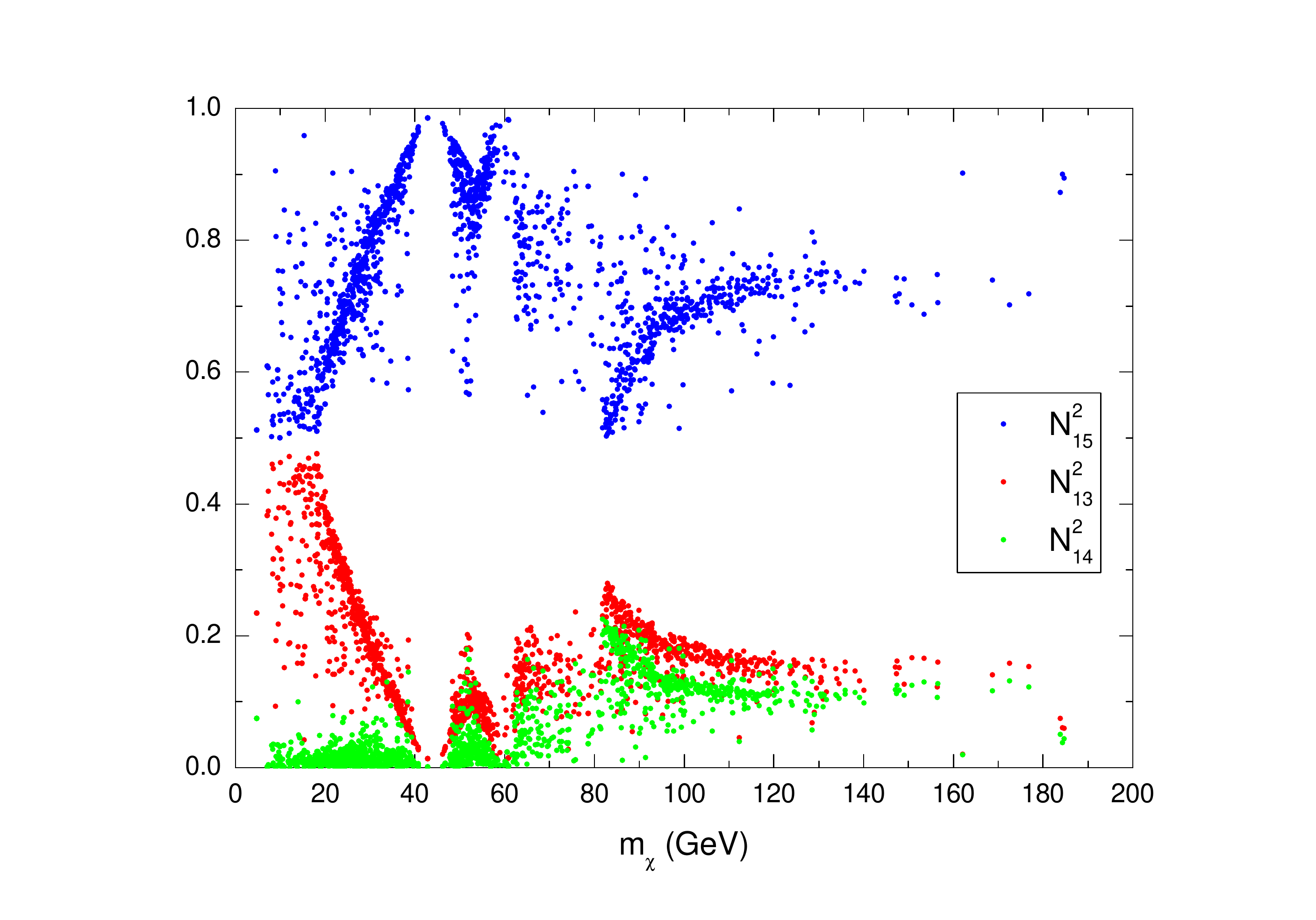}
\includegraphics[width=7cm,height=8cm]{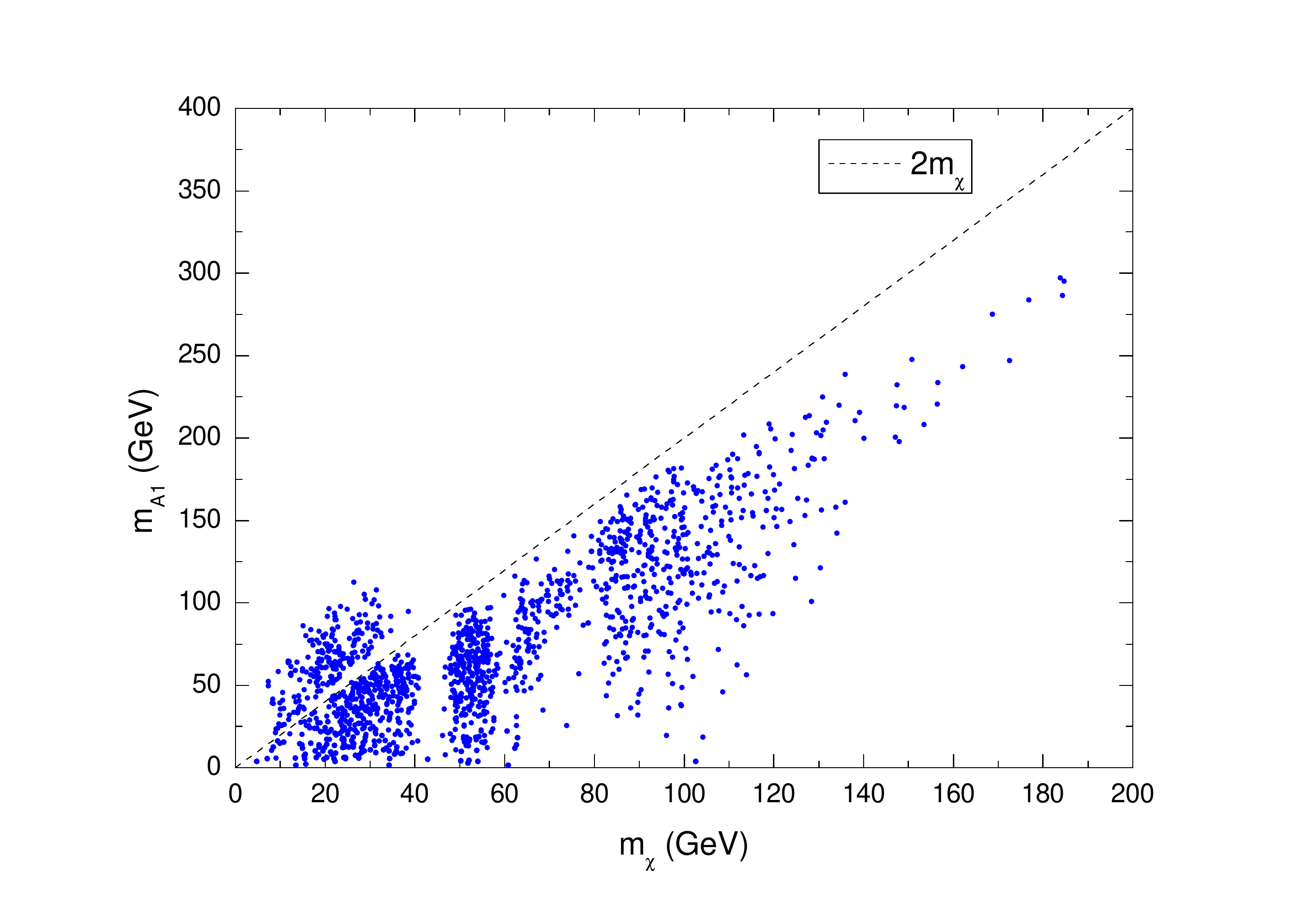}
\centering
 \caption{Samples which satisfy DM relic abundance \cite{Planck}, $N^{2}_{15}\geq 0.5$ and  Higgs mass $m_{h_{2}}=125.1 \pm 1.0$ GeV.
Left plot: the parameter space is projected to  the two-parameter plane of DM mass and $N^{2}_{15}$,
with $\mid N_{13}\mid^{2}$ and $\mid N_{14}\mid^{2}$ shown for illustration.
Right plot: It is projected to two-parameter plane of masses of DM and lighter CP-odd scalar,
where the dotted line refers to the resonant mass relation $m_{A_{1}}=2m_{\tilde{\chi}^{0}_{1}}$.}
\label{Relic}
\end{figure}

Apart from constraints contained in the code,  
there are $2560$ out of $20$ million samples in this scan which satisfy DM relic density and SM Higgs mass constraint simultaneously.
In Fig.\ref{Relic} we show the parameter space in two different ways.
The left plot reveals the dependence of $N_{1i}$ ($i=3-5$) on the DM mass,
which verifies that the singlino component is indeed the main component. 
The right plot indicates the coherence of masses of DM and the lighter CP-odd scalar,
where they trend to saturate at the resonant annihilation $m_{A_{1}}=2 m_{\tilde{\chi}^{0}_{1}}$,
especially for DM mass beneath half of $m_{Z}$. 
The reason is that the coefficient of $T_{A_{1}\tilde{\chi}^{0}_{1}\tilde{\chi}^{0}_{1}}$ and that of mixing angle 
in Eq.(\ref{Annihilation}) are both small in this DM mass region.
A resonant annihilation is therefore needed to compensate the suppression.
But it is modified when the lightest CP-even Higgs $h_1$ makes substantial contribution to the singlino annihilation.

\subsection{Direct Detection}
\begin{figure}
\centering
\includegraphics[width=7cm,height=8cm]{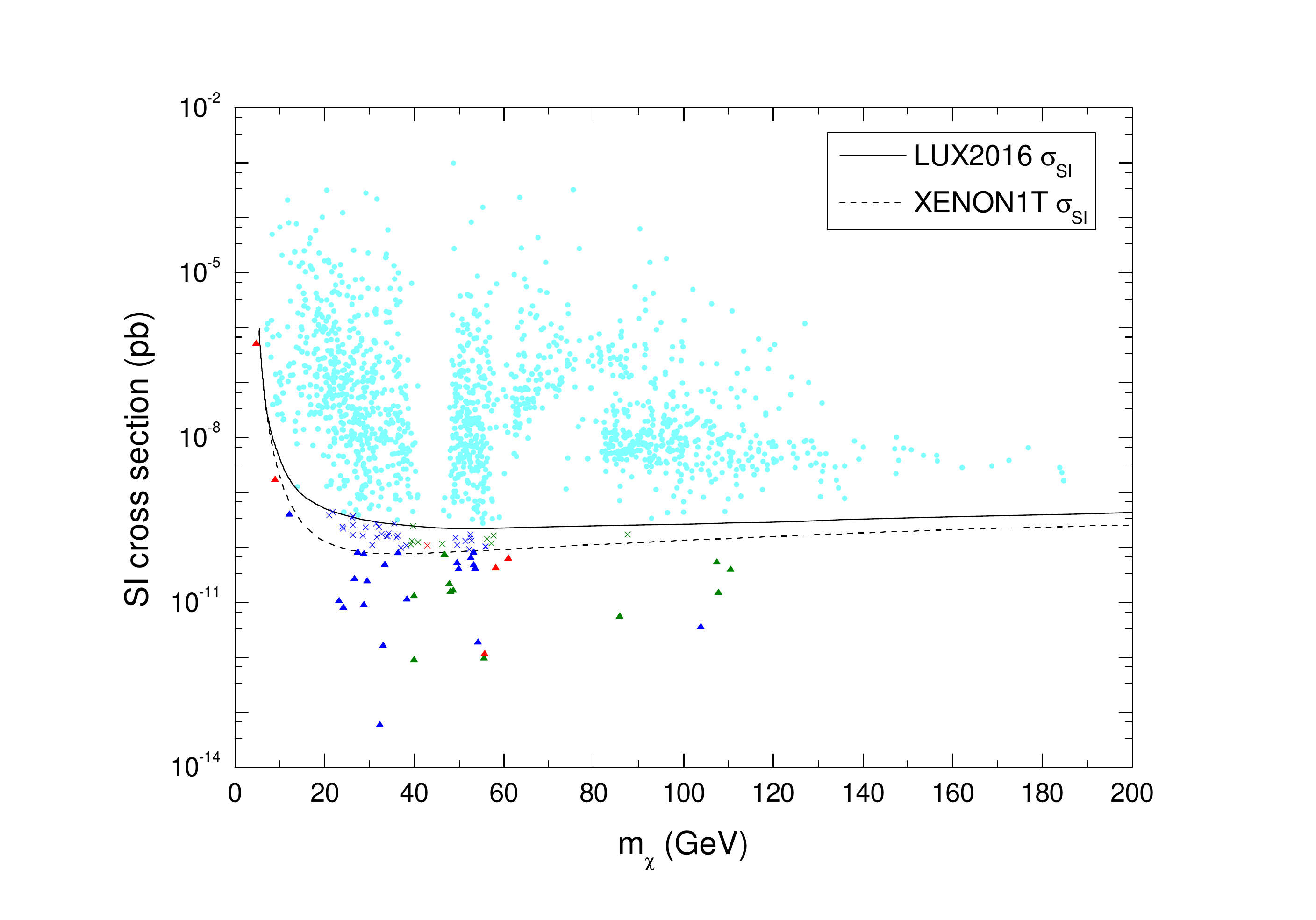}
\includegraphics[width=7cm,height=8cm]{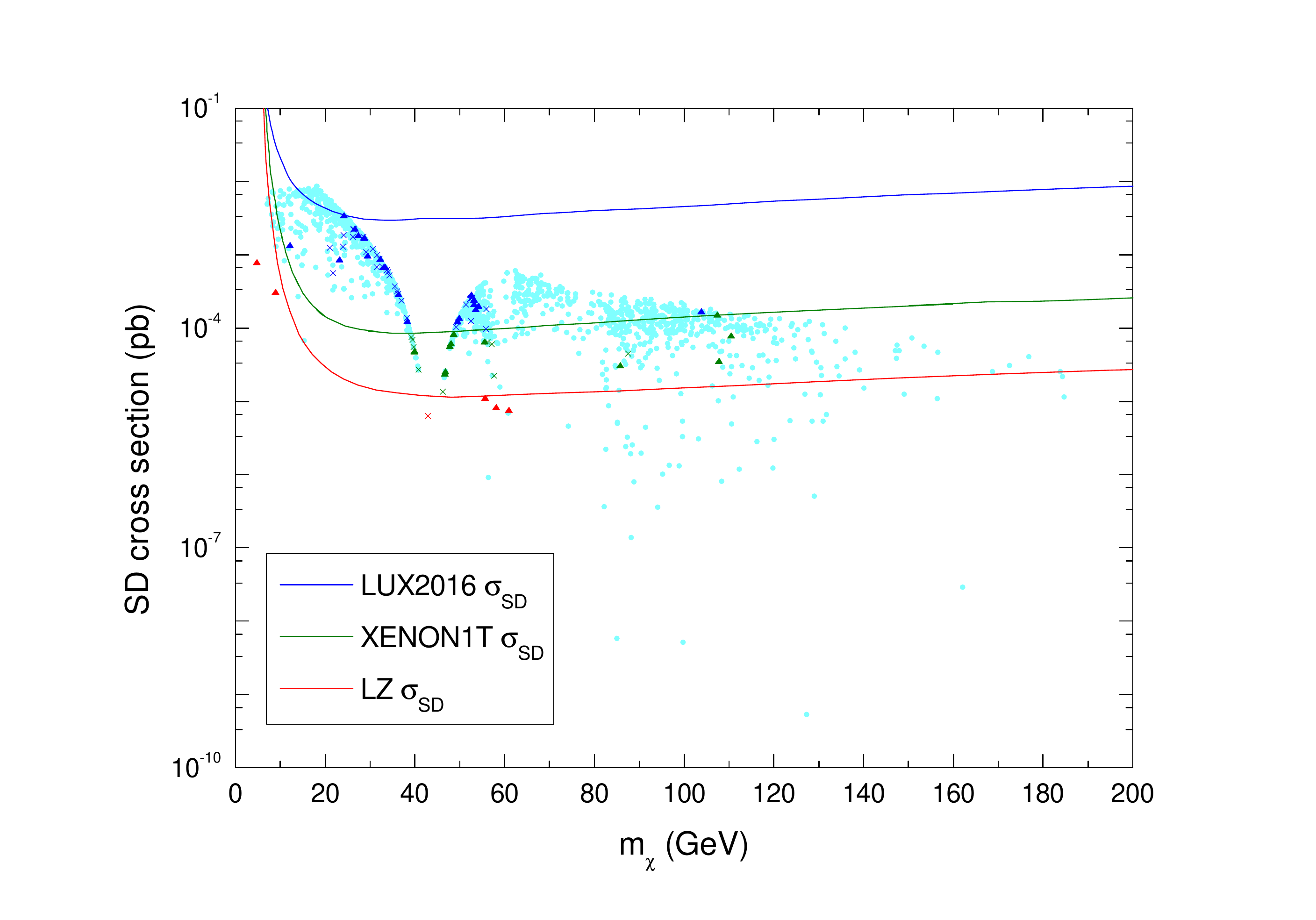}
\centering
 \caption{SI (left) and SD (right) cross section for the same samples of Fig.\ref{Relic}. 
What are referred by $``\bullet"$ and $``\times"$  are excluded by the SI limits at LUX  \cite{LUX2016} and Xenon1T \cite{Xenon1T}, respectively;
and by blue and green are excluded by the SD limit at Xenon1T \cite{XenonSD} and LZ \cite{LZ}, respectively.
In contrast, samples marked by red $``\triangle"$ are not excluded neither by SD or SI limits.}
\label{S}
\end{figure}

Since samples which satisfy the DM relic density have been prepared,
it is straightforward to discuss DM direct detection in terms of DM-nucleon scattering.
The spin-independent (SI) and spin-dependent (SD) scattering cross section $\sigma$
is given by Feynman diagram of interchanging SM Higgs and $Z$ boson, respectively.
By employing micrOMEGAs \cite{1407.6129} we show $\sigma_{\text{SI}}$ and $\sigma_{\text{SD}}$ as function of DM mass $m_{\tilde{\chi}^{0}_{1}}$ in the left and right plot of Fig.\ref{S}, respectively. 
In the left plot samples referred by $``\bullet"$ and by $``\times"$  are excluded by the SI limits at LUX  \cite{LUX2016} and Xenon1T \cite{Xenon1T}, respectively.
Nevertheless, samples referred by $``\triangle"$ are still consistent with these SI limits.
Meanwhile, exclusions by different SD limits are also illustrated by different colors in the left plot,
with blue and green corresponding to Xenon1T \cite{XenonSD} and LZ \cite{LZ}, respectively,
as clearly shown in the right plot.
Therefore, sample referred by red $``\triangle"$ are consistent with both nowadays SI and SD limits,
although the number of them is rare in the figure.

The left plot of Fig.\ref{S} shows that most of samples have been excluded by SI limit at LUX.
The main reason is that $\sigma_{\text{SI}}$ is approximated to be 
\begin{eqnarray}{\label{Sc}}
\sigma_{SI}=\frac{4}{\pi}\left(\frac{m_{\tilde{\chi}^{0}_{1}}m_{p}}{m_{\tilde{\chi}^{0}_{1}}+m_{p}}\right)^{2}f_{p}^{2},
\end{eqnarray}
for moderate or large value of $\tan\beta$. 
Here, $m_{p}$ is the proton mass, and $f_{p}$ is given by
\begin{eqnarray}{\label{f}}
\frac{f_{p}}{m_{p}}\simeq\frac{g}{4M_{W}m^{2}_{h}}T_{h_{2}\tilde{\chi}^{0}_{1}\tilde{\chi}^{0}_{1}}\cdot(f_{Td}-f_{Tu}+f_{Ts}-\frac{2}{27}f_{TG})
\simeq 7\times 10^{-3}\frac{g}{4M_{W}m^{2}_{h}}T_{h_{2}\tilde{\chi}^{0}_{1}\tilde{\chi}^{0}_{1}},
\end{eqnarray}
with
\begin{eqnarray}{\label{Th}}
T_{h_{2}\tilde{\chi}^{0}_{1}\tilde{\chi}^{0}_{1}}\simeq \sqrt{2}\lambda N_{15}(N_{13}\cos\beta+N_{14}\sin\beta).
\end{eqnarray}
The magnitude of $T_{h_{2}\tilde{\chi}^{0}_{1}\tilde{\chi}^{0}_{1}}$ is typically of order $10^{-1}-10^{-2}$.
Substituting Eq.(\ref{Th}) into Eq.(\ref{Sc}) gives rise to $\sigma_{\text{SI}}$ typically of order $10^{-8}$ pb,
which is excluded by the latest LUX 2016 limit \cite{LUX2016}.
$\sigma_{\text{SI}}$ can further decrease when the signs of $N_{13}$ and $N_{14}$ are opposite, 
as a result of which $T_{h_{2}\tilde{\chi}^{0}_{1}\tilde{\chi}^{0}_{1}}$ in Eq.(\ref{Th}) is significantly suppressed 
to help evade the LUX or even Xenon1T limit.

\section{Collider Constraints}
Results in the previous section show that light neutralino DM still survives in the facilities of  LUX,  Xenon1T and LZ.
It is natural to ask what the fate of them is at colliders.
This section is devoted to explore this question.

\begin{figure}
\centering
\includegraphics[width=7cm,height=8cm]{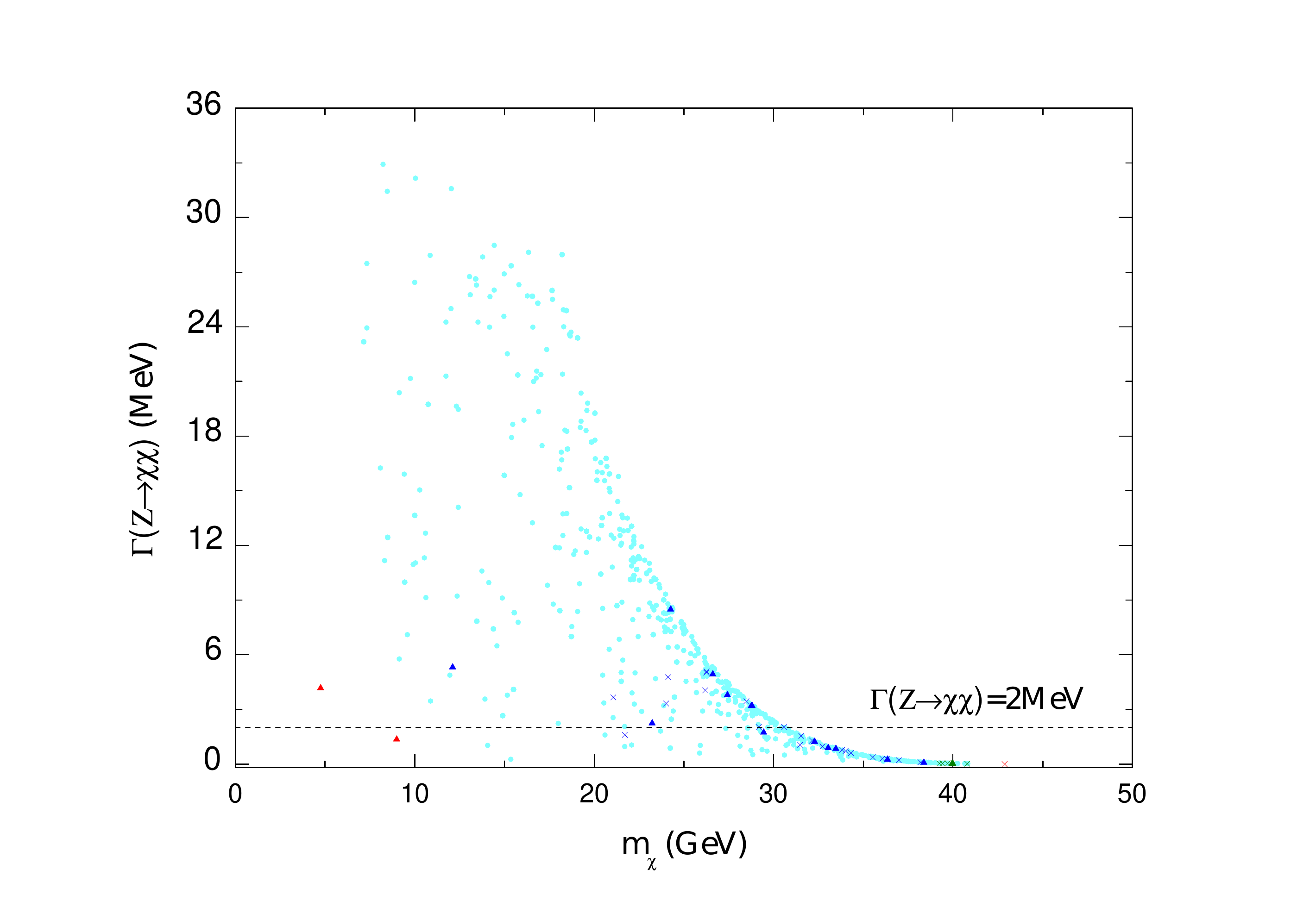}
\includegraphics[width=7cm,height=8cm]{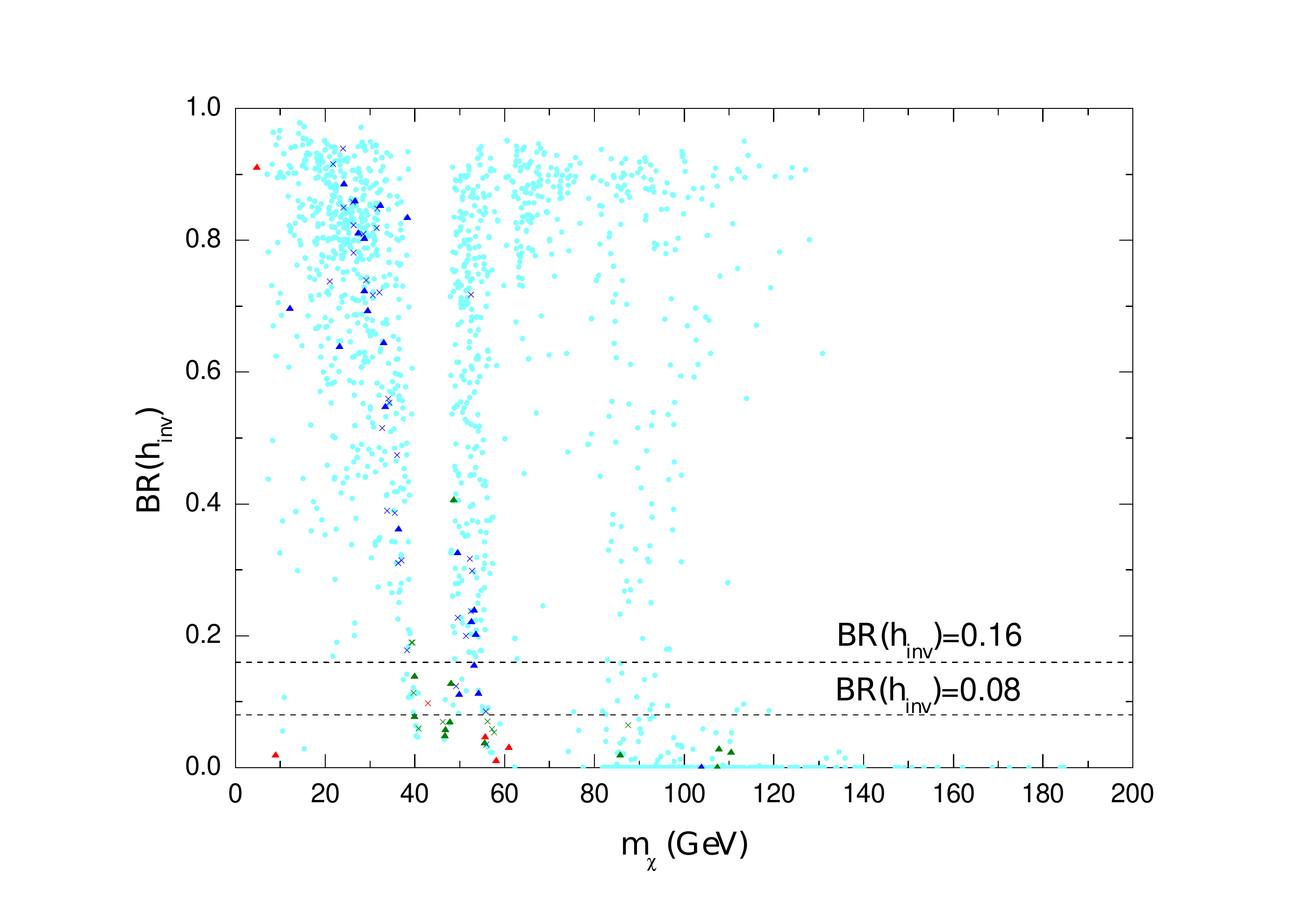}
\centering
 \caption{Left plot: contribution to $Z$ boson invisible decay for the same samples of Fig.\ref{Relic},
where it tells us that the reported upper bound on $\Delta \Gamma_{Z}$ \cite{0509008} has already imposed severely constraint on the parameter space.
Right plot: contribution to SM Higgs invisible decay instead,
where the horizontal dotted line from top to bottom refers to $\Gamma^{inv}_{h}/ \Gamma_{h}=\{0.16,0.08\}$ respectively.
Note that the reference of colors in these two plots are the same as that of Fig.\ref{S}.}
\label{Decay}
\end{figure}

Before we address the constraint imposed by $Z$ boson invisible decay into light neutralino DM,
we postpone the discussion about Higgs invisible decay.
The invisible decay width $\Gamma^{inv}_{Z}$ is determined by \cite{1007.1151}, 
 \begin{eqnarray}{\label{ZDecay}}
\Gamma^{inv}_{Z}=\frac{G_{F}M^{3}_{Z}}{12\sqrt{2}\pi}\left(N^{2}_{13}-N^{2}_{14}\right)^{2}\left(1-\frac{4m^{2}_{\tilde{\chi}^{0}_{1}}}{M^{2}_{Z}}\right)^{3/2}\simeq 0.165~\text{GeV}\cdot\left(N^{2}_{13}-N^{2}_{14}\right)^{2},
\end{eqnarray}
The reported experimental bound $\Gamma^{inv}_{Z}\leq 2$ MeV \cite{0509008} implies that 
the magnitude of $(N^{2}_{13}-N^{2}_{14})^{2}$ in Eq.(\ref{ZDecay}) is upper bounded by $\sim 0.01$ roughly.
It explains why most of samples in Fig.\ref{Relic} are excluded according to the left plot therein.
Fortunately, the left plot of Fig.\ref{Decay} shows that 
a small portion of samples (red triangle) surviving in DM direct detection experiments is still not ruled out by this bound.

Finally, we discuss the constraint arising from SM Higgs invisible decay. 
Since light $A_{1}$ is needed by light singlino-like DM,
the invisible decay width $\Gamma^{inv}_{h}$ may be composed of the following three parts :
 \begin{eqnarray}{\label{HDecay}}
\Gamma(h_{2}\rightarrow h_{1}h_{1})&=&\frac{1}{16\pi m_{h_{2}}}\mid T_{h_{2}- h_{1}-h_{1}}\mid^{2}
\left(1-\frac{4m^{2}_{h_{1}}}{m^{2}_{h_{2}}}\right)^{1/2},\nonumber\\
\Gamma(h_{2}\rightarrow \tilde{\chi}^{0}_{1}\tilde{\chi}^{0}_{1})&=&\frac{1}{8\pi}\mid T_{h_{2}- \tilde{\chi}^{0}_{1}-\tilde{\chi}^{0}_{1}}\mid^{2}
m_{h}\left(1-\frac{4m^{2}_{\tilde{\chi}^{0}_{1}}}{m^{2}_{h_{2}}}\right)^{3/2},\nonumber\\
\Gamma(h_{2}\rightarrow A_{1}A_{1})&=&\frac{1}{16\pi m_{h_{2}}}\mid T_{h_{2}- A_{1}-A_{1}}\mid^{2}
\left(1-\frac{4m^{2}_{A_{1}}}{m^{2}_{h_{2}}}\right)^{1/2}.
\end{eqnarray}
Eq.(\ref{HDecay}) suggests that $\Gamma^{inv}_{h}$ is very sensitive to the Yukawa coefficients $T$s.
Similar to $\Gamma^{inv}_{Z}$ the upper bound $\Gamma^{inv}_{h}\leq 16\%\cdot\Gamma_{h}$ \cite{1606.02266} 
as reported in the LHC Run 1 data has also excluded most of samples.
When DM mass is above half of SM Higgs mass, the second channel in Eq.(\ref{HDecay}) is closed,
but the others may be still allowed. After combing the plots of Fig.\ref{S} we draw the conclusion that 
neutralino DM with mass as light as several GeVs is hardly excluded by the $Z$ boson and Higgs invisible decay.
In Table.\ref{point} we show the main input and out parameters of a benchmark point.

\begin{table}
\tiny{
\begin{center}
\begin{tabular}{|c|c|c|c|c|c|c|c|c|c|c|c|c|c|c|}
\hline
 $\lambda$ & $\kappa$ & $\tan\beta$ & $\mu$  & $M_{3}$  & $m_{\tilde{t}_{3}}$ & $A_{t}$ & $m_{h_{3}}$  
 & $m_{h_{2}}$ & $m_{h_{1}}$ & $m_{A_{2}}$ & $m_{A_{1}}$ & $m_{\tilde{\chi}^{0}_{1}}$ & $\Gamma^{inv}_{Z}$ & $\text{Br}^{inv}_{h}$ \\ 
 \hline
 0.32 & 0.007 & 9.67 & 172.87 & 1708.15 & 1368.64 & -2674.69 & 1687.88 & 125.18& 25.81& 1687.85& 21.09 & 8.98 & 1.35& $1.9\%$   \\
 \hline
\end{tabular}
\caption{Main input parameters and output mass spectrum for a benchmark point in Fig.\ref{Decay}, 
where mass and $\Gamma^{inv}_{Z}$ is in unit of GeV and MeV, respectively. }
\label{point}
\end{center}}
\end{table}

\newpage

\section{Conclusion}
This work presented a numerical study of NMSSM with heavy soft mass parameters associated with the third generation
but  light neutralino DM due to small spontaneously breaking of PQ symmetry.
Naively speaking, such light DM may be already excluded either by DM direct detection facilities or SM $Z$ boson and Higgs scalar invisible decay experiments.
The main finding, however, is that both the latest SD and SI limits from LUX and Xenon1T are unable 
to exclude light neutralino DM with mass of order several GeVs.
This study also demonstrates that even the further precision test on Higgs invisible decay at the HL-LHC 
or LZ experiments fail to exclude such a possibility.

\begin{acknowledgments}
This work is supported in part by the National Natural Science Foundation of China under Grant No.11775039.
\end{acknowledgments}

\appendix
\section{Mass Matrixes}
The mass matrixes discussed in this appendix determine both the DM eigenstate mass 
and its couplings to CP-even Higgs scalar, CP-odd scalar $A_{1}$, $Z$ boson etc.
Firstly, we follow Ref. \cite{0505142} to decompose the Higgs doublet scalars $H^{0}_{u,d}$ and singlet scalar $S$ into the following fields,
\begin{eqnarray}{\label{Hs}}
H^{0}_{u}&=& \upsilon_{u}+\frac{1}{\sqrt{2}}(H_{uR}+iH_{uI}),\nonumber\\
H^{0}_{d}&=& \upsilon_{d}+\frac{1}{\sqrt{2}}(H_{dR}+iH_{dI}),\nonumber\\
S&=& s+\frac{1}{\sqrt{2}}(S_{R}+iS_{I}),
\end{eqnarray}
where  $\upsilon^{2}_{u}+\upsilon^{2}_{d}=(174~\text{GeV})^{2}$ and $s$ is the vacuum expectation value of singlet scalar.
In order to eliminate the Goldstone mode, one rotates the gauge eigenstates from $(H_{uR},H_{dR},S_{R})$ to $(H, h, S_{R})$ via the orthogonal matrix $U(\beta)$ 
with $\tan\beta=\upsilon_{u}/\upsilon_{d}$.
Under the new basis $(H, h, S_{R})$, the CP-even mass squared reads as,
\begin{eqnarray}{\label{CpEven}}
\mathcal{M}^{2}_{S}=
\left(%
\begin{array}{ccc}
\frac{A^{2}_{\lambda}}{1+x}+(M^{2}_{Z}-\lambda^{2}\upsilon^{2})\sin^{2}2\beta & -\frac{1}{2}(M^{2}_{Z}-\lambda^{2}\upsilon^{2})\sin4\beta & -\lambda A_{\lambda}\upsilon\cos2\beta \\
* &   M^{2}_{Z}\cos^{2}2\beta +\lambda^{2}\upsilon^{2}\sin^{2}2\beta &-\lambda A_{\lambda}\upsilon\sin2\beta\frac{x}{1+x}  \\
*& * & \lambda^{2}\upsilon^{2}(1+x) \\
\end{array}%
\right)
\end{eqnarray}
Here $x=m^{2}_{s}/(\lambda^{2}\upsilon^{2})$.
After diagonalization by an orthogonal matrix $S_{ij}$ we obtain three CP-even neutral scalars $h_{i}$,
and one of them is identified as the SM-like Higgs.
The off-diagonal elements in $M^{2}_{S}$ determine the mixing effects between these mass eigenstates,
the magnitude of which are severely upper bounded by the Higgs precision measurements at the LHC \cite{1303.1812}.
With small mixing effects the SM-like Higgs is mainly composed of $h_{2}$.

Under basis $(A,S_{I})$ the mass matrix squared for CP-odd scalars is given by,
\begin{eqnarray}{\label{CpOdd}}
\mathcal{M}^{2}_{P}\simeq
\left(%
\begin{array}{cc}
\frac{2\mu}{\sin2\beta}A_{\lambda} & \lambda\upsilon A_{\lambda}  \\
* &   \frac{\lambda^{2}\upsilon^{2}A_{\lambda}\sin2\beta}{2\mu}   \\
\end{array}%
\right)
\end{eqnarray}
After diagonaliztion by an orthogonal $2\times2$ matrix $P'_{ij}(\theta_{A})$ with angle $\theta_{A}$,
one obtains two CP-odd neutral scalars $A_{i}$ (ordered in mass). 
The determinant of $\mathcal{M}^{2}_{P}$ in Eq.(\ref{CpOdd}) is zero in the PQ limit, 
which implies that there is a massless CP-odd scalar. 
 
On the other hand, the neutralino mass matrix $M_{\chi}$ under the gauge eigenstates $(\tilde{B}, \tilde{W},\tilde{H}_{u},\tilde{H_{d}}, \tilde{s})$ are given by,
\begin{eqnarray}{\label{NeutralinoMass}}
M_{\chi}=\left(
\begin{array}{ccccc}
M_{1} & 0 & M_{Z}s_{W}\sin\beta & -M_{Z}s_{W}\cos\beta & 0\\
 *&   M_{2}  & -M_{Z}c_{W}\sin\beta & M_{Z}c_{W}\cos\beta & 0 \\
* & * & 0 & -\mu &  -\lambda \upsilon_{d} \\
* & * & -\mu & 0 & -\lambda \upsilon_{u} \\
* & * & * & * & \frac{2\kappa}{\lambda}\mu\\
\end{array}%
\right),
\end{eqnarray}
where $\mu=\lambda s$ and $s_{W}=\sin\theta_{W}$.
When $M_{1},M_{2} >>\mu$,  bino and wino are decoupled,
which leads to a singlino-like neutralino LSP.

\end{document}